\documentclass[preprintnumbers,3p,12pt]{elsarticle}

\usepackage{amsmath,amssymb}
\usepackage{mathtools}
\usepackage{bm}
\usepackage{graphicx}
\usepackage{MnSymbol}
\usepackage{url}
\usepackage{hyperref}

\newcommand{\LQ}{\Lambda_{\rm QCD}}

\newcommand{\msbar}{\overline{\rm MS}}

\newcommand{\bea}{\begin{eqnarray}}
\newcommand{\eea}{\end{eqnarray}}
\newcommand{\simgt}{\hbox{ \raise3pt\hbox to 0pt{$>$}\raise-3pt\hbox{$\sim$} }}
\newcommand{\simlt}{\hbox{ \raise3pt\hbox to 0pt{$<$}\raise-3pt\hbox{$\sim$} }}

\newcommand{\be}{\begin{equation}}
\newcommand{\ee}{\end{equation}}

\journal{Physics Letters B}

\begin{document}

\begin{flushright}
    \normalsize TU--1218
\end{flushright}

\begin{frontmatter}

\title{On order $\LQ^2/m$ renormalons in quarkonium system}

\author{Takuya~Agemura and Yukinari~Sumino
\vspace*{3mm}
}

\address{
Department of Physics, Tohoku University,
Sendai, 980--8578 Japan
}%

\begin{abstract}
\small
For the heavy quarkonium system
we examine 
${\cal O}(\LQ^2/m)$ renormalons, which
are expected to be included in the perturbative series of
the pole mass and $1/(mr^2)$ 
interquark potential.
We find indications of existence and cancellation of 
these renormalons,
from examinations of stability and convergence properties of 
the perturbative series and their resummations, as well as
by comparison with the known properties of the ${\cal O}(\LQ)$ renormalons.
\end{abstract}

\end{frontmatter}

\section{Introduction}
\label{sec:Introduction}

The proper treatment of renormalons in theoretical calculations of QCD plays an important role
in recent and near-future high-precision particle physics.
For instance, cancellation of renormalons is incorporated 
in determinations of the fundamental physical constants 
$m_b$, $m_c$, $\alpha_s(M_Z)$, $|V_{cb}|$, etc.
Renormalons have been actively investigated particularly in the context of heavy quarkonium systems.
In fact, the discovery of the cancellation of the ${\cal O}(\LQ)$ renormalons
between 
the twice of the quark pole mass $2m_{\rm pole}$ and the static QCD potential 
$V_{\rm{QCD}}(r)$ \cite{Pineda:id,Hoang:1998nz,Beneke:1998rk}
led to a drastic improvement in the predictability of the observables of the heavy quarkonium systems.
This cancellation is achieved by expressing the pole mass by a short-distance mass such as
the $\msbar$ mass.

In this Letter we focus on the next-to-leading order renormalons
in the heavy quarkonium system, i.e.,
${\cal O}(\LQ^2/m)$ renormalons.
Possible existence of ${\cal O}(\LQ^2/m)$ renormalon in the quark pole mass has been
anticipated and discussed.
Known properties are as follows \cite{Neubert:1996zy}. 
(a) It is induced
by the non-relativistic kinetic energy operator $\bm{D}^2/(2m)$; 
(b) It is not forbidden by any
symmetry, and parametrically it induces an ${\cal O}(\LQ^2/m)$ uncertainty; 
(c) It does not appear in the large-$\beta_0$ approximation.
So far, no strong positive evidence has been reported on the size or effects of
this renormalon.
The next-to-leading order renormalon contained in $V_{\rm{QCD}}(r)$
is of order $\LQ^3 r^2$ \cite{Brambilla:1999qa,Sumino:2020mxk} and no renormalon of order $\LQ^2/m$ is included
(since $V_{\rm{QCD}}(r)$ is independent of $m$).
The counterpart which cancels the ${\cal O}(\LQ^2/m)$ renormalon of $2m_{\rm pole}$
is expected to be included in the potential 
$V_{1/r^2}(r)\sim-1/(mr^2)$, which is a part of the non-relativistic Hamiltonian of the quarkonium system.
In fact, by naive power counting, we expect the size of the renormalon to be
\bea
&&			\delta V_{1/r^2}
			~\sim~ \int_{q\lesssim \LQ}\!\!\!\! d^3\bm{q}\,\,
			\frac{e^{i\bm{q} \cdot \bm{r}}}{m\,q} 
			~\sim~\frac{ {\LQ}^{2}}{m} \,,
\eea
in the case $\LQ \ll 1/r$, where we used the Fourier integral representation of
$V_{1/r^2}(r)$.
Moreover, it has been pointed out that $V_{1/r^2}(r)$ induces anomalously large corrections
to the quarkonium spectrum \cite{Recksiegel:2002za}, which may be a sign of the ${\cal O}(\LQ^2/m)$ renormalon.

\begin{figure}
\begin{center}
\includegraphics[width=7cm]{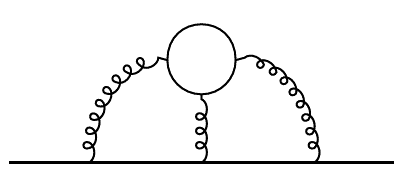}
~~~
\includegraphics[width=5.5cm]{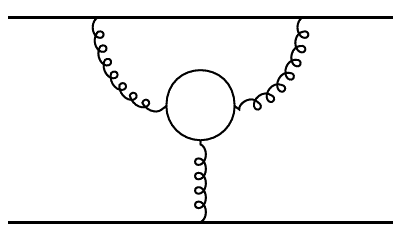}
\end{center}
\caption{Feynman diagrams relevant to the calculations of the 
${\cal O}(\LQ^2/m)$ renormalons of $2m_{\rm pole}$ and $V_{\rm NA}(r)$,
respectively, in the large-$\beta_0$ approximation.
Bubble chain insertions to the gluon propagators of these diagrams need to be computed.}
\label{fig:FeynDiag}
\end{figure}

A frequently-used method to investigate renormalons in perturbative QCD is the
so-called ``large-$\beta_0$ approximation'' \cite{Beneke:1994qe}.
Unfortunately it is difficult to apply this method to the analysis of the 
${\cal O}(\LQ^2/m)$ renormalons of $2m_{\rm pole}$ and $V_{\rm NA}(r)$,
since it would involve difficult calculations of bubble-chain insertions to the diagrams
shown in Fig.~\ref{fig:FeynDiag}.
Instead, we use the recently developed method for estimating renormalons,
the FTRS \cite{Hayashi:2020ylq,Hayashi:2021vdq} and DSRS \cite{Hayashi:2023fgl} methods.\footnote{
FTRS and DSRS stand for ``Renormalon Subtraction by Fourier Transform'' and ``Dual Space  Renormalon Subtraction,'' respectively.
}
These methods utilize the property that the renormalons are suppressed in the
Fourier space or dual space: 
By renormalization-group (RG) improvement of the perturbative series
in these spaces we can separate the renormalons and the rest, and we can
estimate them individually and systematically as more terms of the
perturbative series are included; At all orders they coincide with the conventional 
definition based
on the regularized Borel resummation of the perturbative series.
Furthermore,
for the reason explained below, we restrict the analysis to the 
``maximally non-abelian (MNA) part'' of the perturbative series, corresponding to the part proportional
to $C_F {C_A}^{\! \! n} \alpha_s^{n+1}$.

In Sec.~\ref{sec:2} we examine $V_{1/r^2}$ in the leading-logarithmic (LL) approximation using the
FTRS and DSRS methods.
Secs.~\ref{sec:3} and \ref{sec:4} deal with beyond-the-LL approximation.
In Sec.~\ref{sec:3} we examine the total energy $2 m_{\rm pole} + V_{\rm QCD} + V_{1/r^2}$
of the heavy quarkonium system in the fixed order of the perturbative expansion.
In Sec.~\ref{sec:4} we examine the sizes of the ${\cal O}(\LQ^2/m)$ renormalons
as well as the total energy in the FTRS and DSRS methods.
We give summary and conclusion in Sec.~\ref{sec:5}.

\section{${\cal O}(\LQ^2/m)$ renormalon of $V_{\text{NA}}$ in LL approximation}\label{sec;u=1}
\label{sec:2}

The $1/(mr^2)$ potential in the non-relativistic Hamiltonian of the quarkonium system
is known up to the order $\alpha_s^3$ \cite{Kniehl:2001ju}
within the framework of potential-NRQCD effective field theory
\cite{Brambilla:2004jw}.
This potential appears first at
order $\alpha_s^2$ and its explicit form (in one operator basis) is given by
\bea
&&
V_{1/r^2}=
\frac{\alpha_s^2}{m_{\rm pole}r^2} \Biggl[ 
\frac{C_F (C_F-2 C_A)}{4} +
\frac{C_F \alpha_s}{\pi} 
\Biggl\{ \frac{130 C_A C_F-101C_A^2+(49C_A-8C_F)T_Fn_l}{72} 
\nonumber\\
&&~~~~~~~~
+\frac{\beta_0}{4}  
(C_F-2 C_A) (\gamma_E +\log(\mu r)
)
-\frac{2}{3} C_A
(C_A+2 C_F) (\log 2+2 \gamma_E+2L_{\rm US})
\Biggr\}\Biggr]
\,,
\label{PotentialOrig}
\\&&
\beta_0=\frac{11}{3}C_A-\frac{4}{3}T_F n_l
\,,
~~~~~~~
L_{\rm US}=\log(\mu_{\rm US} r)
\,,
\eea
where the color factors are given by $C_F=4/3$, $C_A=3$ and $T_F=1/2$;
$n_l$ denotes the number of light quark flavors; 
$\alpha_s = \alpha_s(\mu^2)$ denotes the
strong coupling constant 
of the theory with $n_l$ quark flavors only in the $\msbar$ scheme.

Because the Hamiltonian itself is not a direct physical observable,
there arise two problems in analyzing this potential.
One problem is that the $1/(mr^2)$ potential mixes with other operators in the Hamiltonian 
under unitary transformations which do not affect physical observables \cite{Brambilla:1999xj}.\footnote{
Since the leading-order Hamiltonian includes the Coulomb potential
$-C_F\alpha_s/r$, the potential proportional to $(C_F\alpha_s)^2/(mr^2)$ 
changes its coefficient, e.g., by the unitary transformation
with a generator proportional to $i C_F\alpha_s \bm{p}\cdot\bm{x}/(mr)$. 
}
This feature gives a certain ambiguity to the potential to be analyzed for the renormalon cancellation.
The same problem does not occur for the MNA part of the $1/(mr^2)$ potential
(the part proportional to $C_F {C_A}^{\! \! n} \alpha_s^{n+1}$).\footnote{
This is the case when the unitary transformation keeps the leading-order
Hamiltonian unchanged and the generator of the transformation is not enhanced by $C_A$.
There is hardly any reason to consider such an enhanced transformation, since the tree-level quark-antiquark
scattering amplitude does not have a part proportional to $C_A$ in the color-singlet channel.
}
Below, we analyze the MNA part, since this part  remains invariant under unitary transformations
and is expected to reflect most vividly the characteristics of 
the IR renormalons.
We call the MNA part of $V_{1/r^2}(r)$ as the ``non-abelian (NA) potential'' for short
and denote it by $V_{\rm NA}(r)$.

The other problem is the existence of an IR divergence.
$V_{1/r^2}$ or $V_{\rm NA}(r)$ is IR divergent at order $\alpha_s^3$ and beyond.
In the above explicit form, the IR divergence has been subtracted in the $\overline{\rm MS}$ scheme
in momentum space.
Associated to the IR divergence, it has an IR logarithmic term represented by $L_{\rm US}$.
This term is canceled by the UV logarithmic term associated with the UV divergence of
the ultra-soft (US) contribution.
Thus, in a physical observable, $L_{\rm US}$ does not appear but is replaced by logarithms of typical
US scales.
In the case of the perturbative series of the quarkonium spectrum, the physical
logarithmic terms include the 
``QCD Bethe logarithm'' \cite{Kniehl:1999ud} which is analogous to the Bethe logarithm in the Lamb shift.
Up to now, there are no known examples where those US contributions give large effects to physical
observables; rather their contributions are always with moderate size.
Hence, in our analysis below, we either replace $\mu_{\rm US}$ by a typical IR scale and vary it in
a reasonable range, or
replace the US logarithm by 1 or 0 for the sake of an order-of-magnitude estimate.


Now we estimate the size of the ${\cal O}(\LQ^2/m)$ renormalon of the
NA potential using the FTRS method and the 
LL approximation.
Then we 
demonstrate that  the convergence of the perturbative series of the NA potential
improves by subtraction of the renormalon.

The FTRS method utilizes the fact that IR renormalons are suppressed by 
Fourier transformation.
As a result, in momentum space convergence is expected to be good.
For this reason,
it would make sense to approximate the NA potential in momentum space by its
LL approximation.
Thus, 
we set
	\begin{equation}
		\begin{aligned}[b]
			\tilde{V}_{\text{NA}}^{\text{LL}}(q)
			&=-\frac{\pi^{2}C_{A}C_{F}}{m_{\text{pole}}q}    {\alpha_{s}(q^2)}^{2}\\
			&\simeq -\frac{\pi^{2}C_{A}C_{F}}{m\, q}     
			      \sum_{n=0}^{\infty}(n+1)\left[\frac{\beta_{0}}{4\pi}
             \log\biggl(\frac{\mu^{2}}{q^{2}}\biggr)\right]^{n}  \alpha_{s}(\mu^{2})^{n+2} \,,
		\end{aligned}
	\end{equation}
where (only in this section) $\alpha_s(q^2)=4\pi/[\beta_0 \log(q^2/\Lambda_{\msbar}^2)]$
denotes the one-loop running coupling constant and we set $n_l=0$ to take the MNA part.
Within our approximation it does not make a difference whether we use the pole mass $m_{\rm pole}$
or the $\msbar$ mass $m\equiv m_{\overline{\rm MS}}(m_{\overline{\rm MS}})$, and we use the latter
in the second line. 
The NA potential in position space is obtained by inverse Fourier transformation
of each term of the series expansion in $\alpha_{s}(\mu^{2})$:
\begin{equation}
	\begin{aligned}[b]
		V_{\text{NA}}^{\text{LL}}&=
		\sum_{n=0}^{\infty}C_{n+1}\alpha_{s}(\mu^{2})^{n+2} \,,
	\end{aligned}
\label{defVLLNA}
 \end{equation}
where
\begin{equation}
	\begin{aligned}[b]
		C_{n+1}&=-\frac{\pi^{2}C_{A}C_{F}}{m}{\biggl(\frac{\beta_{0}}{4\pi}\biggr)}^{n}
		\int \frac{d^3\bm{q}}{(2\pi)^3} \frac{e^{i\bm{q} \cdot \bm{r}}}{q} 
		    (n+1)  \left[\log\left(\frac{\mu^2}{q^2}\right)\right]^n.\\
	\end{aligned}
\label{defCn}
\end{equation}
The Borel transform of $V_{\text{NA}}^{\text{LL}}$ is given by
\bea
&&
		B_{V_{\rm NA}^{\rm LL}}(u)=\sum_{n=0}^{\infty} \frac{C_{n+1}}{(n+1)!}\left(\frac{4\pi u}{\beta_{0}}\right)^{n+1}
		=-\frac{4\pi^{3}C_{A}C_{F}}{m\beta_{0}}u
		\int \frac{d^3\bm{q}}{(2\pi)^3} \frac{e^{i\bm{q} \cdot \bm{r}}}{q}
		\left(\frac{\mu^{2}}{q^{2}}\right)^{u}
\nonumber\\ && ~~~~~~~~~
		=-\frac{2\pi^{3/2}C_{A}C_{F}}{m\beta_{0}r^{2}}\, \left(\frac{\mu^{2}r^{2}}{4}
\right)^{u}\hspace{3pt}\cfrac{u \,\Gamma(1-u)}{\Gamma(\frac{1}{2}+u)}.
		\label{BVNA}
\eea
Thus, there are poles at $u=1,2,3,\cdots$ in the Borel plane. 
The standard prescription for the Borel resummation is to regularize the 
inverse Borel transform by deforming the integral contour of the Borel integral
into the upper or lower half $u$-plane.
The size of the ${\cal O}(\LQ^2/m)$ renormalon is defined as the imaginary part of
the regularized Borel integral around $u=1$.
Hence, it can be calculated from the residue of the pole at $u=1$ as
\begin{equation}
	\begin{aligned}[b]
		[\delta V_{\text{NA}}^{\text{LL}}]_{u=1}^{\mathrm{Borel}}
		&=\frac{2\pi}{i\beta_{0}} \oint_{-C_{1}}du
\left(-\frac{2\pi^{3/2}C_{A}C_{F}}{m\beta_{0}r^{2}}\right)\left(\frac{\mu^{2}r^{2}}{4}\right)^{u}
		\hspace{3pt} \cfrac{\Gamma(2-u)}{\Gamma(\frac{1}{2}+u)} \frac{u}{1-u}e^{-\tfrac{4\pi}{\beta_{0}\alpha_{s}}u}\\
		&=-\frac{4\pi^{3}C_{A}C_{F}}{{\beta_{0}}^{2}} \frac{{\Lambda_{\overline{\mathrm{MS}}}}^{2}}{m}.
	\label{eq;LLB}
	\end{aligned}
\end{equation}
Thus, the $u=1$ renormalon of the NA potential is independent of $r$.


In the FTRS method, there is a shortcut to calculate the same quantity
(regularized Borel resummation) bypassing the Borel transform.
%
In the inverse Fourier transformation, we deform the integral path to circumvent the pole of $\alpha_{s}(q^2)$. 
Then the renormalon can be extracted as the imaginary part of the regularized inverse Fourier integral.
Thus,
	\bea
	&&		[\delta V_{\text{NA}}^{\text{LL}}]^{\text{FTRS}}=
			-\frac{\pi C_{A}C_{F}}{2mr} \oint_{C}\frac{dq}{2\pi i} \sin (qr)\alpha_{s}(q^2)^{2}\nonumber\\
		&&~~~~~~~~~~~~~~~~
	=-\frac{\pi C_{A}C_{F}}{2mr} \oint_{C}\frac{dq}{2\pi i}\left [qr-\frac{1}{6}(qr)^{3}+\cdots\right]\alpha_{s}(q^2)^{2}\,,
\\&&
			[\delta V_{\text{NA}}^{\text{LL}}]^{\text{FTRS}}_{u=1}=
			-\frac{\pi C_{A}C_{F}}{2m} \oint_{C}\frac{dq}{2\pi i} \,q\,\alpha_{s}(q^2)^{2}
			=-\frac{4\pi^{3}C_{A}C_{F}}{{\beta_{0}}^{2}} \frac{{\Lambda_{\overline{\mathrm{MS}}}}^{2}}{m}\,,
		\label{eq;LLF}
	\eea
where $C$ is the loop surrounding the pole of $\alpha_{s}(q^2)$ at $q=\Lambda_{\msbar}$.
The linear term in the expansion of $\sin(qr)$ is taken to obtain the $u=1$ renormalon. 
The result in eq.~\eqref{eq;LLF} indeed matches the calculation by the Borel resummation method in 
eq.~\eqref{eq;LLB}.

\begin{figure}[htbp]
\centering
\includegraphics[keepaspectratio, width=7.5cm]{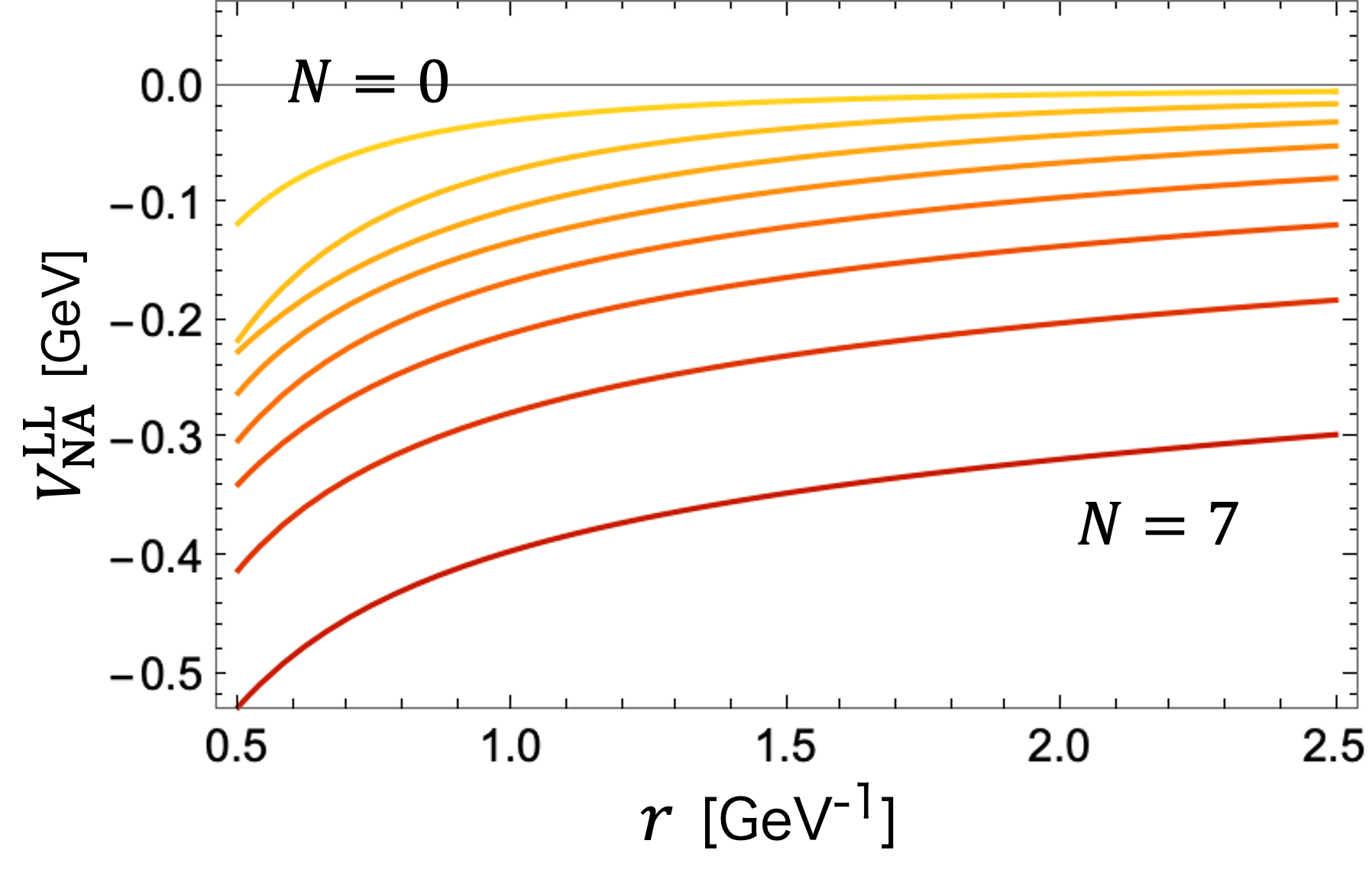}
~~~
\includegraphics[keepaspectratio, width=7.9cm]{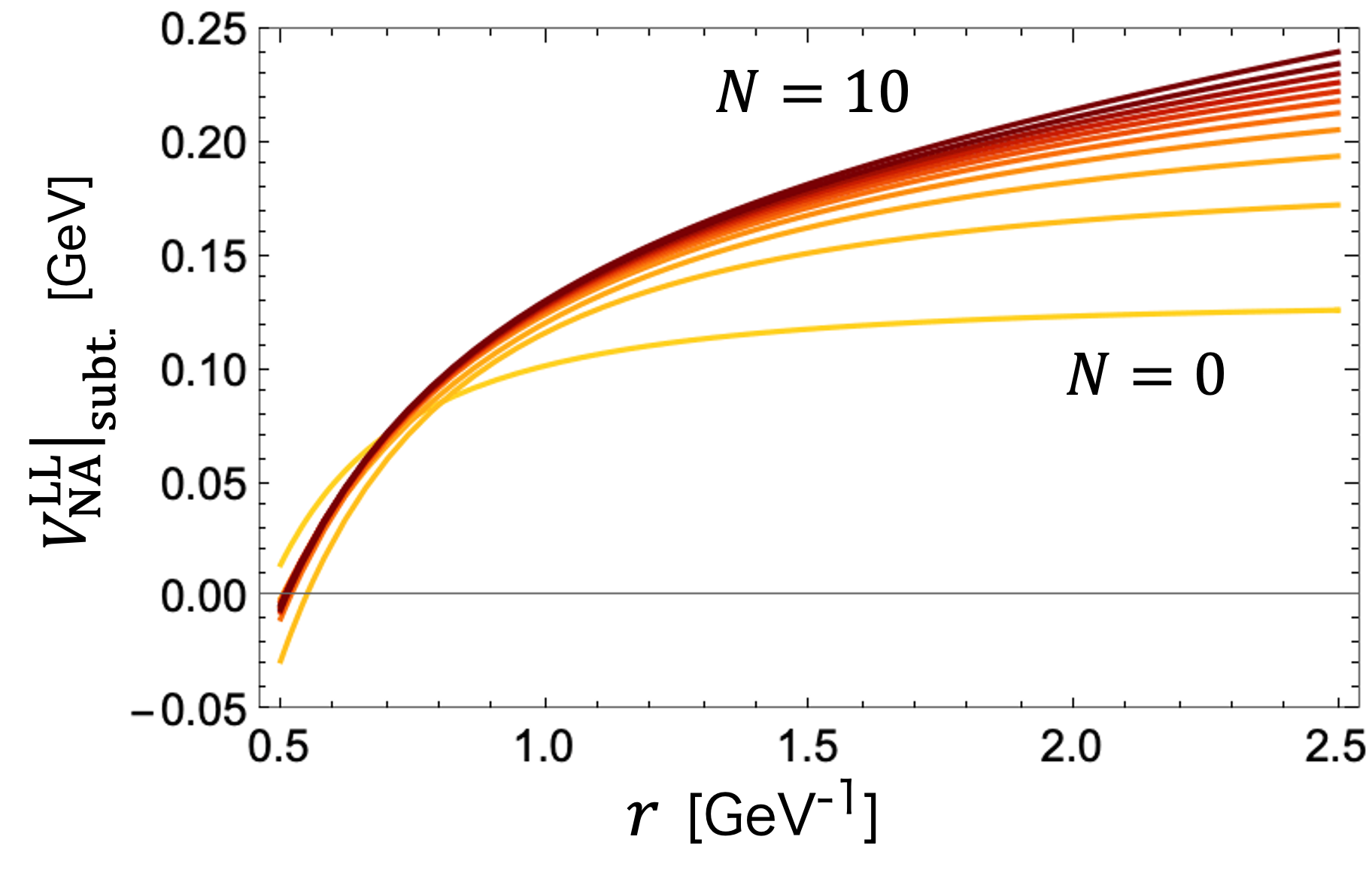}
{$~~~~~~~~$(a)}~~~~~~~~~~~~~~~~~~~~~~~~~~~~~~~~~~~~~~~~~~~~~~~~~~~~~~~~~~
{(b)$~$}
\caption{(a) NA potential in the LL approximation by the FTRS method,
expanded in $\alpha_s(\mu)$ and truncated at order ${\alpha_{s}^{N+2}}$.
(b) The same potential after subtraction of the $u=1$ renormalon contributions.
Note that the scale of the vertical axis in (b) is magnified by a factor of two
compared to (a).}
\label{Fig:VLL}
\end{figure}

The asymptotic behavior of the perturbative series induced by the $u=1$ renormalon can be extracted
from the pole at $u=1$ of the Borel transform in eq. (\ref{BVNA}) as
\bea
B_{V_{\rm NA}^{\rm LL}}(u) \sim -\frac{2\pi^{3/2}C_{A}C_{F}}{m\beta_{0}r^{2}} \, \frac{\mu^{2}r^{2}}{4}\,\cfrac{1}{\Gamma({3}/{2})}\,\frac{u}{1-u}\
=-\frac{\pi C_{A}C_{F}}{m\beta_{0}}\mu^{2}\sum_{n=0}^{\infty}u^{n+1} .
\eea
Therefore, the order ${\alpha_{s}^{n+2}}$ term of $V^{\text{LL}}_{\text{NA}}$ behaves asymptotically as
\begin{equation}
\begin{aligned}[b]
V_{\text{NA}}^{\text{LL},(n)}&\sim
-\frac{C_{A}C_{F}{\alpha_{s}}^{2}}{4m}\mu^{2}\left(\frac{\beta_{0}\alpha_{s}}{4\pi}\right)^{n}(n+1)!.
\label{const}
\end{aligned}
\end{equation}
This $r$-independent (constant-potential-like) behavior represents the contribution from the $u=1$ renormalon. 
Fig.~\ref{Fig:VLL}(a) shows the truncated perturbative expansion of the NA potential in the LL approximation
[eq.~\eqref{defVLLNA}], and 
Fig.~\ref{Fig:VLL}(b) shows the potential after subtracting the renormalon contribution eq.~\eqref{const}. 
Each line represents the NA potential up to order ${\alpha_{s}^{N+2}}$, with the input parameters $\mu=3$ GeV, 
$\Lambda_{\msbar}=0.298$~GeV,\footnote{
This corresponds to 
$\alpha_s(2\,{\rm GeV})=0.3$,
which we use throughout this Letter as a reference effective coupling close to that of the
bottomonium states.
}
$n_{l}=0$, and $m=4.19$~GeV
[$\alpha_s(\mu^2)=0.247$].
In Fig.~\ref{Fig:VLL}(a), as the perturbative order increases, the potential shifts downwards
almost by a constant independent of $r$, indicating a lack of convergence by the renormalon contributions
[cf., eq.~\eqref{const}]. 
On the other hand, in Fig.~\ref{Fig:VLL}(b), after subtracting the $u=1$
renormalon contributions, the convergence is notably improved.
This demonstrates that removing the $u=1$ renormalon enhances the convergence of the perturbative expansion.
These features are qualitatively similar to the behaviors of the QCD potential before
and after subtraction of the ${\cal O}(\LQ)$ ($u=1/2$) renormalon. (See Figs.~11 of Ref.~\cite{Sumino:2014qpa}.)

Next, we use the DSRS method to calculate the renormalon of
$V_{\text{NA}}^{\text{LL}}(r)$ defined by eqs.~\eqref{defVLLNA}
and \eqref{defCn}.
While at all orders the DSRS method gives the same result as the
regularized Borel resummation or the FTRS method, 
its systematic improvement is different from the FTRS method
as higher-order terms
are included. 
We choose the parameters of the DSRS method as $(a,u')=(1,-1)$ which suppress the
renormalons at $u=1,2,3,\cdots$ in the dual space.
Following the prescription of Ref.~\cite{Hayashi:2023fgl}, we obtain
	\begin{equation}
		\begin{aligned}[b]
			V_{\text{NA}}^{\text{LL}}(r)
			&=\int_{0}^\infty dp^{2}\,e^{-r^{2}p^{2}}\,\,\overline{V}_{\text{NA}}^{\text{LL}}(p^{2})
\,,
\label{DSRSrep-VNALL}
		\end{aligned}
	\end{equation}
where the NA potential in the dual space is given by
\bea
&&
\overline{V}_{\text{NA}}^{\text{LL}}(p^{2}) =
-\frac{2\pi C_{A}C_{F}}{\beta_{0}m}\sum_{n=0}^{\infty}(n+1)!
\,\tilde{a}_{n+1}\,\left(\frac{\beta_{0}}{4\pi}\right)^{n+1}\alpha_{s}(p^{2})^{n+2}
\,,
\label{dualspVLLNA}
\\ &&
\sum_{n=0}^{\infty}\tilde{a}_{n+1}u^{n+1}=\frac{u\,\Gamma(1+u)}{\Gamma(1+2u)}
\,.
\label{def-an}
\eea
After truncating the series expansion in eq.~\eqref{dualspVLLNA} at an arbitrary order,
we deform the integral path of eq.~\eqref{DSRSrep-VNALL} into the upper or lower-half
complex $p^2$ plane to circumvent the pole of $\alpha_s(p^2)$.
Similarly to the FTRS method, the size of the $u=1$ renormalon is obtained as
	\begin{equation}
		\begin{aligned}[b]
			[\delta V^{\text{LL}}_{\text{NA}}]^{\text{DSRS}}_{u=1}&=-\frac{8\pi^{3}C_{A}C_{F}}{{\beta_{0}}^{2}m}
			\sum_{n=0}^{\infty}\tilde{a}_{n+1}(n+1)! \, \oint_{C}\frac{dp^{2}}{2\pi i}\,\frac{1}{[\log(p^{2}/{\Lambda_{\overline{\mathrm{MS}}}}^{2})]^{n+2}}\\
			&=-\frac{8\pi^{3}C_{A}C_{F}}{{\beta_{0}}^{2}} \frac{{\Lambda_{\overline{\mathrm{MS}}}}^{2}}{m} \sum_{n=0}^{\infty}\tilde{a}_{n+1}.
		\label{renoVLLD}
		\end{aligned}
	\end{equation}
The sum over $n$ converges.
By setting $u=1$ in eq.~\eqref{def-an}, we find $\sum_{n=0}^{\infty}\tilde{a}_{n+1}=\frac{1}{2}$ and the previous
result is reproduced.

\begin{figure}[tbh]
\centering
\includegraphics[keepaspectratio, width=7.5cm]{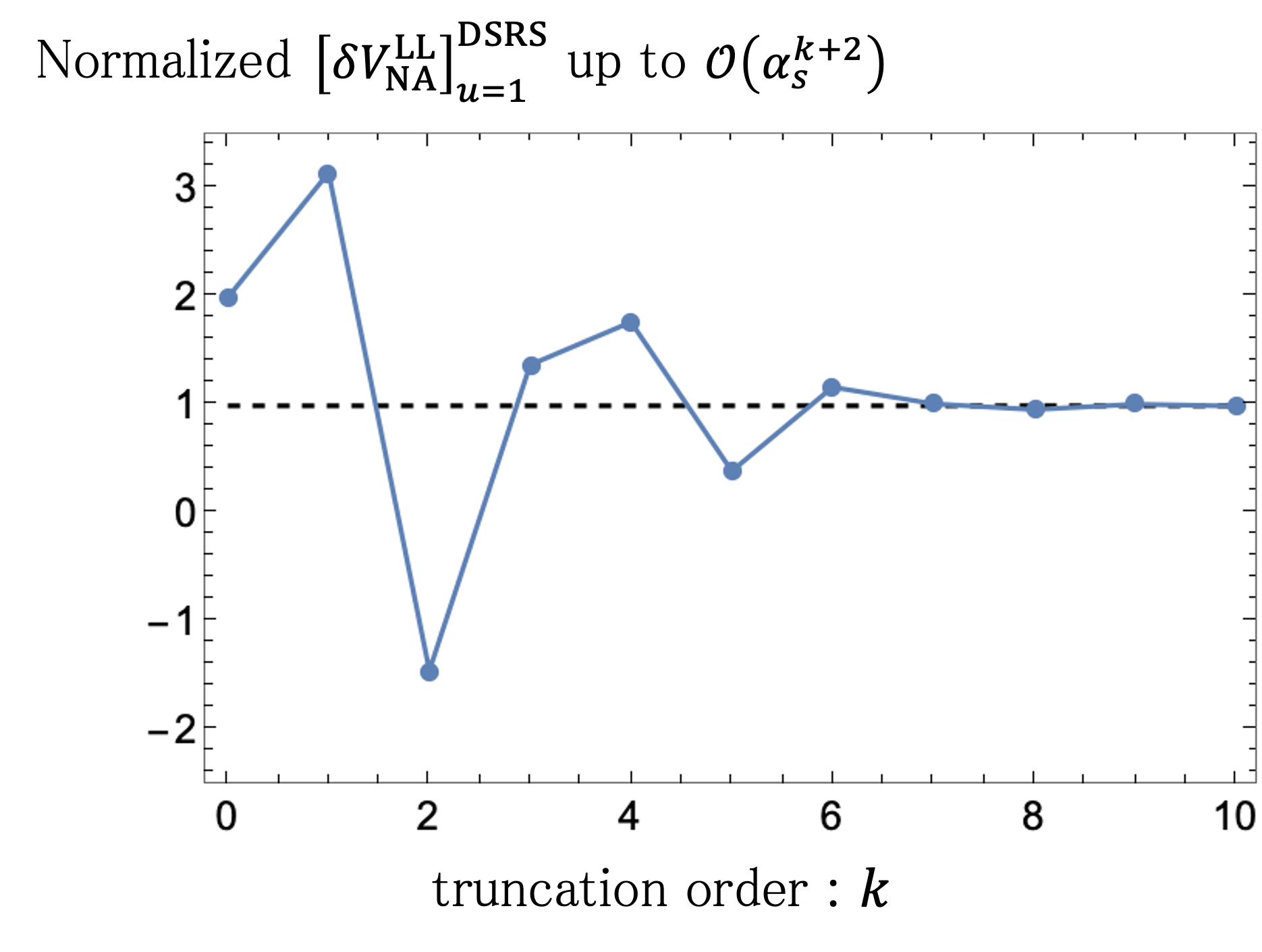}
~~~
\includegraphics[keepaspectratio, width=7.9cm]{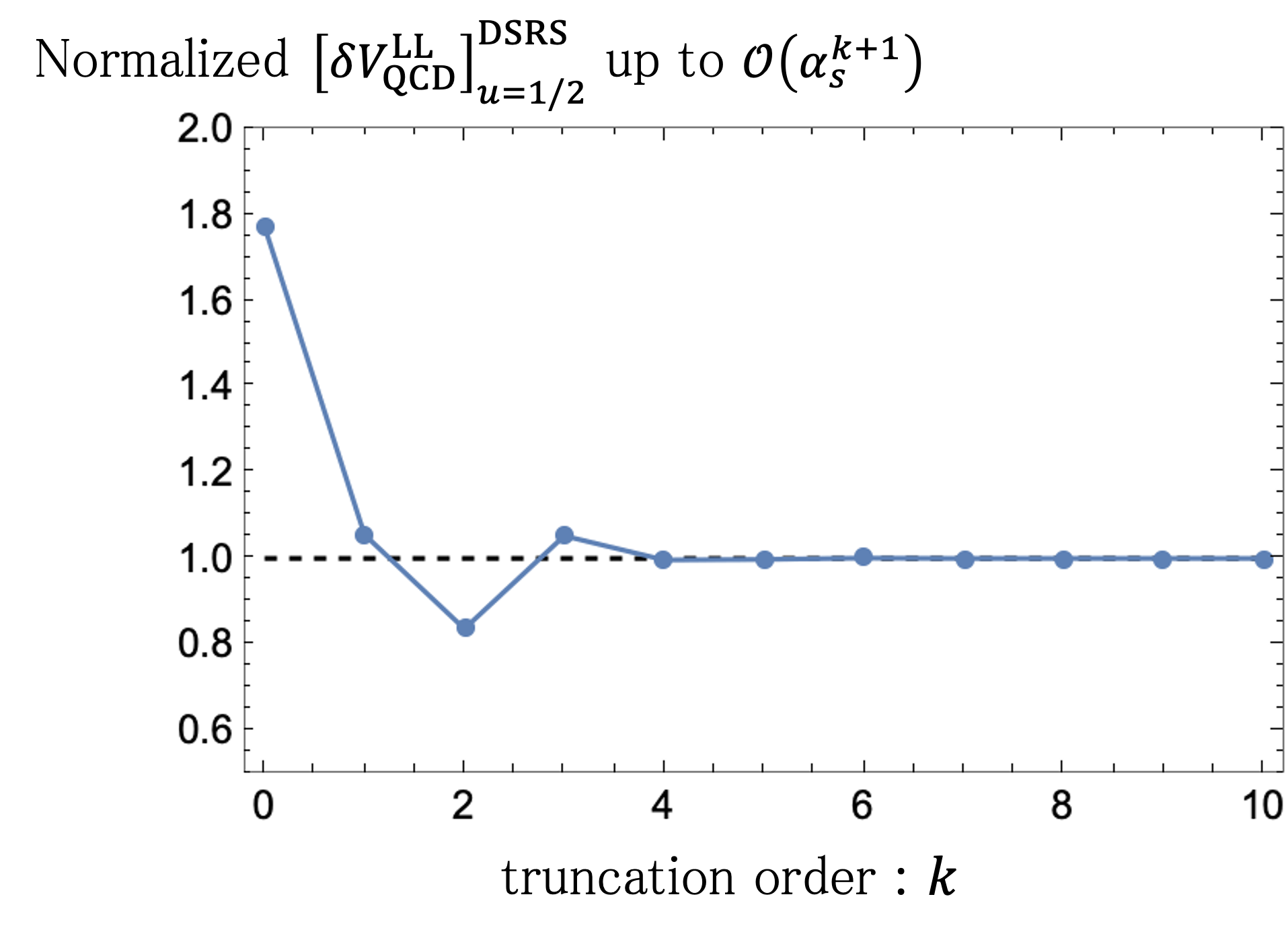}
{$~~~~~~~~$(a)}~~~~~~~~~~~~~~~~~~~~~~~~~~~~~~~~~~~~~~~~~~~~~~~~~~~~~~~~~~
{(b)$~$}
\caption{(a) Size of the $u=1$ renormalon of $V_{\rm NA}^{\rm LL}$ in the DSRS method
as we vary the truncation order.
(b)~Size of the $u=1/2$ renormalon of $V_{\rm QCD}^{\rm LL}$ in the DSRS method
as we vary the truncation order.
Both series are normalized by the exact values.
Note that the scale of the vertical axis in (b) is magnified by a factor of 4
compared to (a).}
\label{fig:BD}
\end{figure}

Let us examine how the size of the renormalon obtained by the DSRS method converges towards the result of the Borel resummation or the FTRS method. 
Fig.~\ref{fig:BD}(a) compares the values obtained by truncating the summation in eq.~(\ref{renoVLLD}) at $n=k$ with the $u=1$ renormalon calculated by the FTRS method (normalized to unity).
At low orders, the truncated sum oscillates rather rapidly.
As the order increases, the sum converges towards the value of FTRS slowly.
From the order $n=7$ the truncated sum starts to approximate the FTRS value with better than 10 per cent accuracy.
For comparison, Fig.~\ref{fig:BD}(b) shows the similar calculation of
the $u=1/2$ renormalon in the QCD potential in the LL approximation using the DSRS method. 
The size of the $u=1/2$ renormalon is also normalized to unity. 
Note that the vertical-axis scales differ by a factor of 4 between the two figures.
It can be observed that the convergence speed of the $u=1$ renormalon in $V^{\text{LL}}_{\text{NA}}$ is 
considerably slower compared to the $u=1/2$ renormalon of the QCD potential.
It can be shown that the expansion coefficients of the $u=u_*$ renormalon
in the DSRS method includes the factor $u_*^n$ as $n$ is varied.
Hence, in general the convergence tends to become worse as the renormalon is
located further away from the origin in the Borel plane.

We have also examined the real part of the resummed potential $V_{\rm NA}^{\rm LL}$ in the 
FTRS and DSRS methods.
The FTRS method gives the Borel resummed potential in the principal-value (PV) prescription and the DSRS method
gives a series which converges to it, with our current construction of the potential.
The former potential approximates well the $N=7$ truncated potential in Fig.~\ref{Fig:VLL}(b)
(up to an additive constant) in the displayed range, which corresponds to the most convergent line
in this figure (truncation at the minimal term).
This feature is consistent with the similar property observed for the QCD potential \cite{Sumino:2003yp}.
For the DSRS potential, the convergence is fairly slow in the displayed range.
The convergence is better at smaller $r$.
More explicitly,
evidently converging behavior can be seen only at $r\lesssim 1~{\rm GeV}^{-1}$ and for $k \geq 8$.
We have to calculate up to higher $k$ to observe a sign of convergence at larger $r$.
On the other hand, we observe a better convergence at smaller distances $r < 0.5~{\rm GeV}^{-1}$.


It is difficult to perform a similar analysis for the $u=1$ renormalon of the pole mass.
This is because the $u=1$ renormalon has a subleading structure $\alpha_s^2\times[\beta_0\alpha_s\log(\mu /m)]^n$,
which resides in between the more dominant leading logarithmic structure 
$\alpha_s\times[\beta_0\alpha_s\log(\mu /m)]^n$, and it is difficult to extract the $u=1$ renormalon as clearly as
the NA potential.
Instead we will
perform a more involved analysis in Sec.~\ref{sec:4} to assess the $u=1$ renormalon of the pole mass.

\section{Beyond LL approximation: Fixed-order analysis of $2m_{\rm pole}$, $V_{\rm QCD}$ and $V_{\rm NA}$}
\label{sec:3}

In this section we examine the perturbative series of the
total energy of the quarkonium system,
$E_{\text{tot}}=2m_{\rm pole}+V_{\rm QCD}+V_{\rm NA}$, 
and discuss cancellation of renormalons among
$2m_{\rm pole}$, $V_{\rm QCD}$ and $V_{\rm NA}$.
We examine the MNA part of their exact perturbative expansions in $\alpha_s(\mu^2)$ up to order $\alpha_s^3$
(fixed-order analysis).

Formally at all orders of perturbative expansion, the dependence of an observable 
on the renormalization scale $\mu$ disappears. 
Therefore, if the perturbative series converges, $\mu$ dependence decreases 
as the order of the perturbative expansion increases.
This is not necessarily the case if the perturbative expansion shows poor convergence. 
In this way, convergence and stability with respect to the scale $\mu$  of the perturbative expansion are closely related. 
Below, we examine the $\mu$ dependence of the total energy $E_{\text{tot}}$ and consider the effects of renormalon cancellations.

It is believed that, in the quarkonium system, the $u=1/2$ renormalons cancel 
between $2m_{\rm pole}$ and $V_{\rm QCD}$.
We consider it plausible that also the $u=1$ renormalons cancel out in the MNA part of the entire expression of
$E_{\text{tot}}=2m_{\text{pole}}+V_{\text{QCD}}+V_{\text{NA}}$. 
Since the $u=1$ renormalon is subdominant, to observe the effects of the cancellation of this renormalon, 
it is necessary to cancel the dominant $u=1/2$ renormalons. 
To clarify the role of each renormalon contribution, we compare the following three cases:\footnote{
Exclusion of $V_{\rm NA}$ in $E_{{1/r}}$ is to achieve cancellation of the  ${\cal O}(\LQ)$ renormalon but not the ${\cal O}(\LQ^2/m)$ renormalon, only by expressing $m_{\rm pole}$ by the $\overline{\rm MS}$ mass.
The role of $V_{\rm NA}$ in the analysis below may be analogous to the role of $V_{\rm QCD}$ in a comparison between $2m_{\rm pole}(m)$ and $2m_{\rm pole}(m)+V_{\rm QCD}(r)$ at a fixed $r$, where stability against scale variation and convergence improve in the latter case. 
}
(I)~$E_{\text{pole}}=2m_{\rm pole}+V_{\rm QCD}+V_{\rm NA}$ without rewriting $m_{\rm pole}$, 
(II)~$E_{{1/r}}=2m_{\rm pole}+V_{\rm QCD}$ after rewriting $m_{\rm pole}$ by the $\msbar$ mass $m$,
(III)~$E_{\text{full}}=2m_{\rm pole}+V_{\rm QCD}+V_{\rm NA}$ after rewriting $m_{\rm pole}$ by the $\msbar$ mass $m$.
It is understood that the MNA part is taken, and in addition we also retain the order $\alpha_s^0$
term ($2m_{\rm pole}$ in $E_{\text{pole}}$ or $2m$ in $E_{1/r}$, $E_{\text{full}}$) for convenience.
We expect that the following renormalons remain uncanceled in each case:
(I) ${\cal O}(\LQ)$ and ${\cal O}(\LQ^2/m)$,
(II) ${\cal O}(\LQ^2/m)$, (III) None, up to $u=1$.

In the cases (II) and (III) we use the MNA part of the pole mass expressed by the $\msbar$ mass
\cite{Melnikov:2000qh},
\bea
&&
m_{\rm pole}^{\rm MNA}=m\,\Biggl[
1 + \frac{C_F\alpha_s(m^2)}{\pi}
\biggl\{1 + \frac{C_A\alpha_s(m^2)}{\pi}
\left(
\frac{1111}{384}-\frac{\pi^2}{12}+\frac{\pi^2}{4}\log 2-\frac{3}{8}\zeta(3)
\right)
\nonumber\\&&
~~~~~~~~~~~~~~
+\Bigl( \frac{C_A\alpha_s(m^2)}{\pi} \Bigr)^2\biggl(
\frac{1322545}{124416}+\frac{1955\pi^2}{3456}+\frac{179\pi^4}{3456}+
\frac{115\pi^2}{72}\log 2 -\frac{11\pi^2}{36}\log^2 2 
\nonumber\\&&
~~~~~~~~~~~~~~ ~~~~~~~~~~~
- \frac{11}{72}\log^4 2
-\frac{11}{3}\rm{Li}_4\Bigl(\frac{1}{2}\Bigr)-\frac{1343}{288}\zeta(3)-\frac{51\pi^2}{64}\zeta(3)+
\frac{65}{32}\zeta(5)
\biggr)\biggr\}\Biggr]
\,.
\eea
We re-emphasize that $\alpha_s$ denotes the 
coupling constant 
of the theory with $n_l$ quark flavors only.
For $V_{\rm NA}$ we use the MNA part of eq.~\eqref{PotentialOrig}.
The MNA part of the QCD potential is given by \cite{Schroder:1998vy}
\bea
&&
V_{\rm QCD}^{\rm MNA}=-\frac{C_F\alpha_s(r^{-2})}{r}
\Biggl[ 1+ \frac{C_A\alpha_s(r^{-2})}{\pi}
\left(\frac{31}{36}+\frac{11}{6}\gamma_E\right)
\nonumber\\&&
~~~~~~~~~~~~~~
+\Bigl( \frac{C_A\alpha_s(r^{-2})}{\pi} \Bigr)^2\biggl(
\frac{4343}{2592}+\frac{247}{54}\gamma_E+\frac{121}{36}\gamma_E^2+\frac{229\pi^2}{432}
-\frac{\pi^4}{64}+\frac{11}{24}\zeta(3)
\biggr)\Biggr]
\,.
\eea
We rewrite $\alpha_s(m^2)$ and $\alpha_s(r^{-2})$ by $\alpha_s(\mu^2)$ using the two-loop relation
and re-expand the
perturbative series by $\alpha_s(\mu^2)$,
in which we set $n_l=0$ to maintain consistency with the MNA part.

\begin{figure}[tbp]
\centering
\includegraphics[keepaspectratio, width=7.5cm]{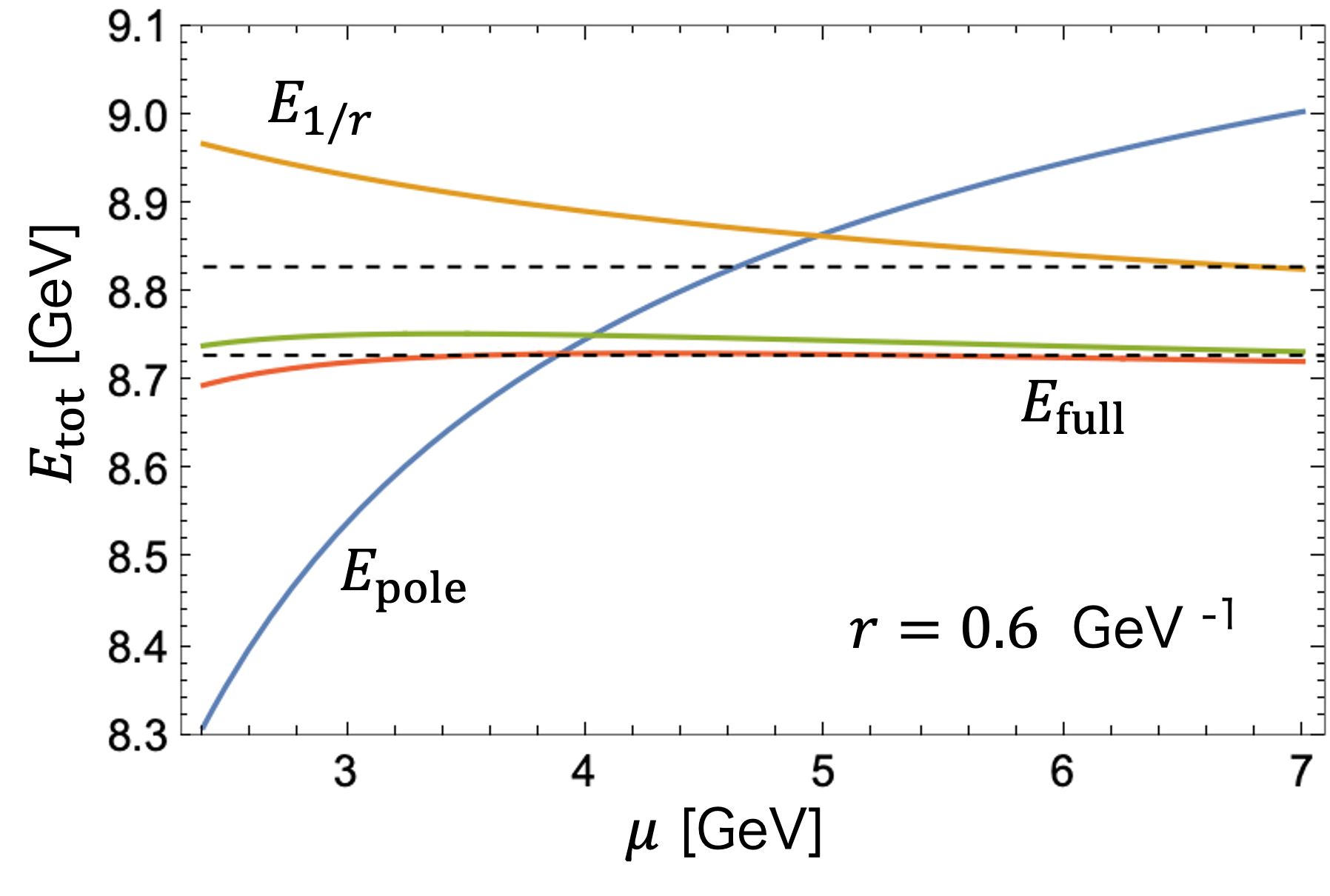}
~~~
\includegraphics[keepaspectratio, width=7.5cm]{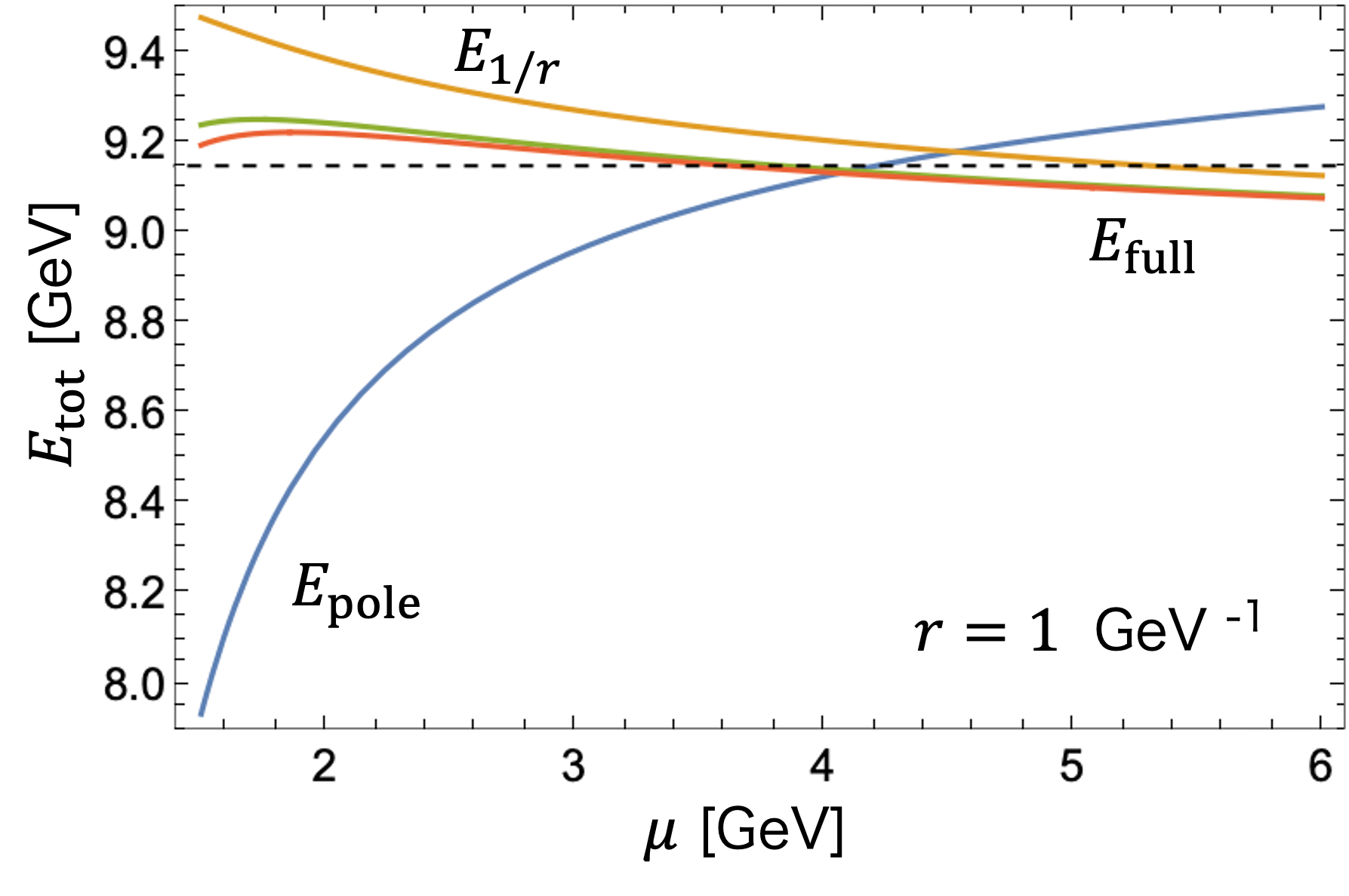}
{$~~~~~~~~$(a)}~~~~~~~~~~~~~~~~~~~~~~~~~~~~~~~~~~~~~~~~~~~~~~~~~~~~~~~~~~
{(b)$~$}
\caption{
The total energies $E_{\text{pole}}$, $E_{1/r}$ and $E_{\text{full}}$
vs.\ $\mu$ at
(a) $r=0.6~\text{GeV}^{-1}$,
(b) $r=1~\text{GeV}^{-1}$.
The green (red) line represents $E_{\text{full}}$ with $\mu_{\rm US}=0.5~{\rm GeV}$ ($1~{\rm GeV}$).
Horizontal dashed lines are shown as a guide.
}
\label{Fig:mu-dep}
\end{figure}

Using the above expressions, we compare the $\mu$ dependences of $E_{\text{pole}}$, $E_{1/r}$ and $E_{\text{full}}$
 in
Figs.~\ref{Fig:mu-dep}(a) and (b).
The input parameters are taken as 
$\alpha_s(2\,{\rm GeV})=0.3$, $m_{\rm pole}=5$~GeV in $E_{\text{pole}}$, and $m=4.19$~GeV in $E_{1/r}$ and $E_{\text{full}}$.
We vary $\mu_{\rm US}=0.5$~GeV and $1$~GeV in $E_{\text{full}}$.
The interquark distance $r$ is fixed
at $0.6~\text{GeV}^{-1}$ and $1~\text{GeV}^{-1}$ in
Figs.~\ref{Fig:mu-dep}(a) and (b), respectively.
We observe that the $\mu$ dependence decreases in the order $E_{\text{pole}}$, $E_{1/r}$ and $E_{\text{full}}$. 
This is reasonable with regard to the expectation that the $u=1/2$ and $u=1$ renormalons get canceled 
in this order.
This order of the $\mu$ dependence of the three total energies is unchanged for 
$0.5~{\rm GeV}^{-1}\lesssim r \lesssim 2.5~{\rm GeV}^{-1}$.
At shorter distances, however, the order of $\mu$ dependence between 
$E_{1/r}$ and $E_{\text{full}}$ is reversed and
the $\mu$ dependence of $E_{1/r}$ becomes the smallest.
These features agree with the following consideration.
The effects of the renormalons are expected to be more prominent at IR, so that the ordering of
the $\mu$ dependence according to the expected residual renormalons would be most naturally realized at larger $r$.
On the other hand, at short distances $r\ll\Lambda_{\msbar}^{-1}$,
$V_{\rm QCD}$ and $V_{\rm NA}$
are mostly perturbative.
Hence, $V_{\rm NA}\propto 1/r^2$ is most
enhanced and its $\mu$ dependence is also magnified at the same time.
This explains why at shorter distances the order of $\mu$ dependence between 
$E_{1/r}$ and $E_{\text{full}}$ is reversed.

Let us fix $\mu$ to the scale where $E_{\text{full}}$ becomes least sensitive
to $\mu$ (the minimal sensitivity scale) and compare the perturbative series of
the three total energies.
For $r=0.6~{\rm GeV}^{-1}$, $dE_{\text{full}}/d\mu$ vanishes at $\mu=4.28$~GeV, where
the perturbative series are given by
\bea
		&&E_{\text{pole}}
		=10.00-0.447-0.358-0.423 \hspace{10pt} \text{GeV}\,, \label{eq;Etotpole}\\
		&&E_{1/r}\,\,
		=8.380+0.268+0.150-0.085 \hspace{10pt}\text{GeV}\,, \\
		&&E_{\rm full}\,\,
		=8.380+0.268+0.096-0.012 \hspace{10pt}\text{GeV}\,. \label{eq;EtotMS}
\eea
We set $\mu_{\rm US}=1$~GeV in $E_{\text{pole}}$ and $E_{\text{full}}$.
Thus, the convergence becomes better in the order $E_{\text{pole}}$, $E_{1/r}$, $E_{\text{full}}$,
as expected.
We also confirm that the $\mu$ dependence of $E_{\text{full}}$ is reduced in a healthy manner
as more terms of the perturbative expansion are included.
The qualitative features are the same also at $r=1~{\rm GeV}^{-1}$ and/or with the choice $\mu_{\rm US}=0.5$~GeV.

\section{Beyond LL approximation: FTRS and DSRS analysis}
\label{sec:4}

In this section we 
estimate the sizes of the $u=1$ renormalons of $V_{\mathrm{NA}}$ and $2\, m_{\rm pole}^{\rm MNA}$
using the FTRS and DSRS methods with their exact perturbative series
up to order $\alpha_s^3$.
We also examine the renormalon-subtracted part of $V_{\rm QCD}^{\rm MNA}(r)+V_{\rm NA}(r)$ in the FTRS and
DSRS methods in the range of $r$ of the bottomonium size.

Let us briefly explain the FTRS method.
For a dimensionless observable $X$ with a mass scale $Q$, we take the 
Fourier transform as
\bea
\tilde{X}(\tau)=\int d^3 \bm{x}\,e^{-i\bm{\tau}\,\cdot\,\bm{x}} \, \rho^{2au'} \, X(\rho^{-a})
\,,
~~~
Q=\rho^{-a}, ~~~ |\bm{x}|=\rho, ~~~ |\bm{\tau}|=\tau 
\,.
\eea
By adjusting the parameters $(a,u')$, we can suppress 
a series of renormalons at $u_*=-u'+n/a$ for $n=0,1,2,\cdots$
of the perturbative series of $\tilde{X}(\tau)$
(defined as singularities in the Borel plane).
Taking advantage of the improved convergence we perform RG improvement
of the perturbative series of $\tilde{X}(\tau)$.
The original observable is restored by inverse Fourier transformation as
\bea
X(Q=\rho^{-a})_\pm^{\rm FTRS} = \frac{\rho^{-2au'-1}}{2\pi^2}
\int_{C_\mp} d\tau\,\tau\,\sin(\tau\rho)\,\tilde{X}(\tau)
\,.
\eea
We can regularize and
separate the series of IR renormalons 
by deforming the integral contour infinitesimally below or above the real $\tau$-axis.
By raising accuracy of the RG-improved series for $\tilde{X}(\tau)$ as LL, NLL, NNLL, $\dots$,
we can estimate the individual renormalons (imaginary part)
and the renormalon-subtracted part (real part) systematically.
The DSRS method is constructed by a similar procedure using inverse Laplace (dual) transformation.

\begin{table}[t]
		\centering
		\begin{tabular}{cc|ccc|c}
		\hline
			&	& $\delta V_{\mathrm{NA}}$ & $2\,\delta m_{\rm pole}^{\rm MNA}$ & $\delta V_{\mathrm{NA}}+2\,\delta m_{\rm pole}^{\rm MNA}$ & $R$\\
		\hline
		{FTRS}	& ${\cal O}(\alpha_{s}^{2})$ & $0.0091$ & $-0.0045$ & $~\,0.0046$ & $0.66$\\
			& ${\cal O}(\alpha_{s}^{3})$ & $0.0062/0.0164$ & $-0.0243$ & $-0.0182/-0.0079$ & $0.41/0.81$\\
		\hline
		{DSRS}& ${\cal O}(\alpha_{s}^{2})$ & $0.1091$ & $-0.1005$ & $\,~0.0086$ & $0.96$\\
			& ${\cal O}(\alpha_{s}^{3})$ & $0.172/0.294$ & $-0.0667$ & $~\,0.105/0.228$ & $0.56/0.37$\\
		\hline
		\end{tabular}
\caption{\label{tab:Renormalon_cancellation_2}
Estimates of the
$u=1$ renormalons in GeV units at each order of the expansions in the
FTRS and DSRS methods. 
The input parameters are $m=4.19$~GeV and $\alpha_s(2\,{\rm GeV})=0.3$.
The parameters of the 
FTRS and DSRS methods are chosen as $(a,u')=(2,-1/2)$.
For ${\cal O}(\alpha_{s}^{3})$ of $\delta V_{\mathrm{NA}}$, $\log(\mu_{\rm US}r)+\gamma_E$ is set to one or zero.
$R$
represents the level of cancellation of the $u=1$ renormalons [eq.~\eqref{Level-Cancel}].
}
	\end{table}

We 
estimate the sizes of the $u=1$ renormalons of $V_{\mathrm{NA}}$ and $2\, m_{\rm pole}^{\rm MNA}$
using the FTRS and DSRS methods.
We use the order $\alpha_s(q^2)^k$ and  $\alpha_s(p^2)^k$ ($k=2,3$)
 expressions of the perturbative series
(with three-loop running coupling constant) in the Fourier space and dual space, respectively.
We choose the parameters of the 
FTRS and DSRS methods as $(a,u')=(2,-1/2)$ which suppress the renormalons at
$u=1/2, 1, 3/2, 2, \cdots$ in the Fourier space and dual space, since
the pole mass includes the $u=1/2$ renormalon as well.
The result is summarized in Tab.~\ref{tab:Renormalon_cancellation_2}.
For estimating the US effects,  
$\log(\mu_{\rm US}r)+\gamma_E$ is set to one or zero
before taking the Fourier transform or dual transform.
The last column shows
\bea
R=1-\frac{|\delta V_{\mathrm{NA}}+2\,\delta m_{\rm pole}^{\rm MNA}|}{|\delta V_{\mathrm{NA}}|+|2\,\delta m_{\rm pole}^{\rm MNA}|}\,,
\label{Level-Cancel}
\eea
which
represents the level of cancellation of the $u=1$ renormalons between
$V_{\mathrm{NA}}$ and $2\, m_{\rm pole}^{\rm MNA}$.
Because we can use only the first two terms for $V_{\rm NA}$ and because the convergence is
expected to be slow (see Sec.~\ref{sec:2}), we expect to observe only a limited signal of renormalons and their
cancellation.
In fact, in both methods we hardly see any sign of convergence for the
estimates of the sizes of the renormalons.
Moreover, the results are fairly dependent on our treatment of the US effects of $V_{\rm NA}$.
On the other hand, we see a tendency of partial cancellation between
$\delta V_{\mathrm{NA}}$ and $2\delta m_{\rm pole}^{\rm MNA}$
at each order.
This tendency is expected in the case that the convergence improves by the cancellation.\footnote{
In Tab.~\ref{tab:Renormalon_cancellation_2}
we choose common $(a,u')$ for $\delta V_{\mathrm{NA}}$ and $2\, \delta m_{\rm pole}^{\rm MNA}$,
because only in that case partial cancellation at each order is expected.
}
(The high cancellation rate for the DSRS order $\alpha_s^2$ case is presumably accidental,
considering the size of each term and the convergence behavior in the LL approximation.)

\begin{figure}[tbp]
\centering
\includegraphics[keepaspectratio, width=10cm]{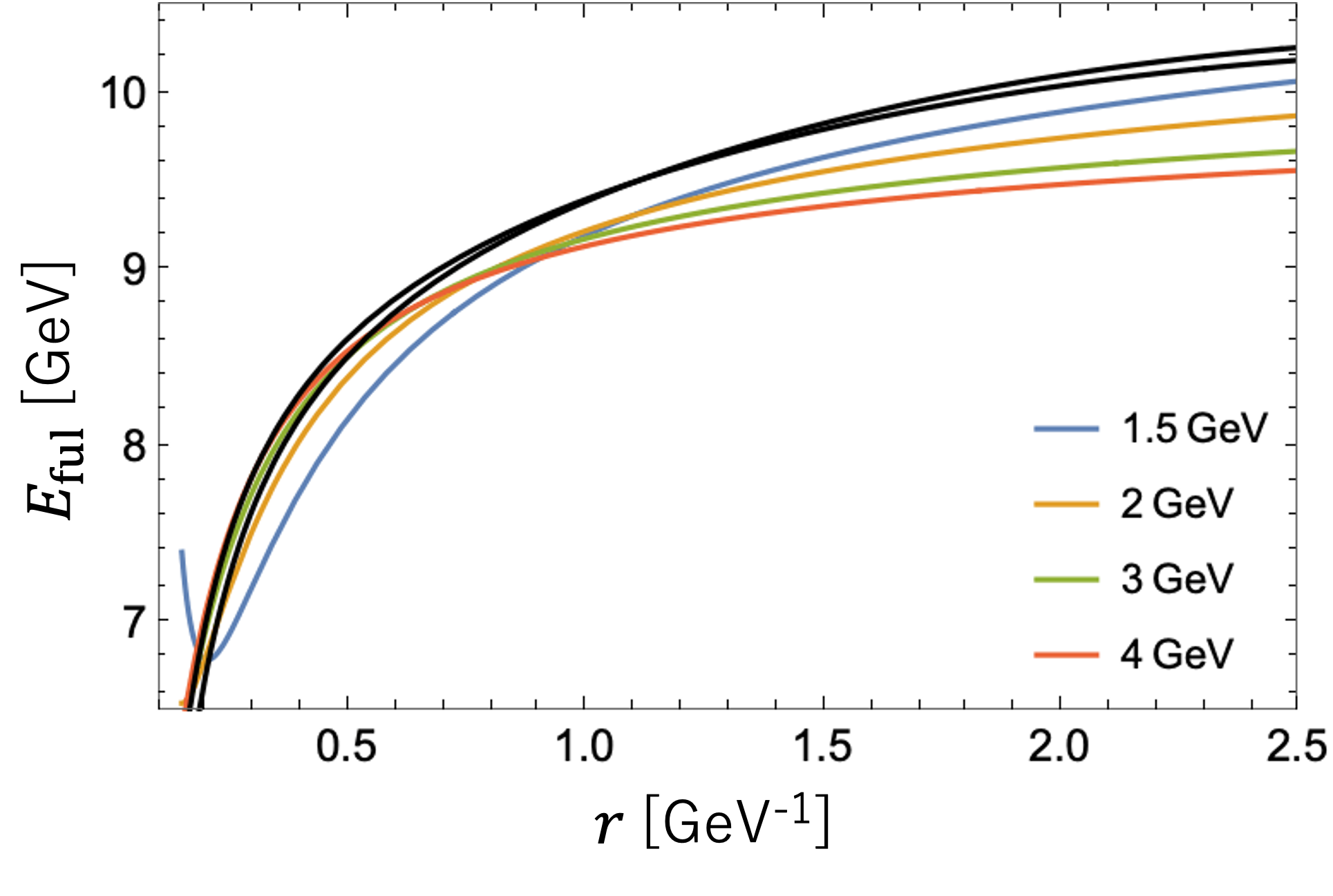}
\caption{
Comparison of
$V_{\rm QCD}^{\rm MNA}+V_{\rm NA}$ in the FTRS method (black lines) and
$E_{\text{full}}$ in the fixed-order perturbative series (colored lines)
as functions of $r$.
An arbitrary $r$-independent constant is added to the former.
The latter lines are for $\mu=4,3,2,1.5$~GeV from left to right.
In the former potential $\log(\mu_{\rm US}r)+\gamma_E$ is set to one or zero
and the result changes only slightly.
In the latter $\mu_{\rm US}$ is set to 1~GeV.
}
\label{Fig:FTRS-vs-FixedOrder2}
\end{figure}

Next we calculate the real part (PV part) of $V_{\rm QCD}^{\rm MNA}+V_{\rm NA}$ 
in the FTRS method and
compare it with the fixed-order total energy $E_{\rm full}$ calculated in the previous section.
Since 
by looking into small $r$ region we can detect UV region compared to the renormalons (typically at $\LQ$ scale),
we expect a better convergence behavior than above.
The comparison is shown in Fig.~\ref{Fig:FTRS-vs-FixedOrder2}.
We choose $(a,u')=(1,-1/2)$ for $V_{\rm QCD}^{\rm MNA}$ and $(a,u')=(1,-1)$ for $V_{\rm NA}$
such that the renormalons of the individual potentials are suppressed minimally.
(These correspond to the ordinary Fourier transformation of both potentials.)
We add an $r$-independent constant to the FTRS result
to facilitate the comparison.
For estimating the US effects,  
$\log(\mu_{\rm US}r)+\gamma_E$ is set to one or zero
before the Fourier transform, by which $V_{\rm QCD}^{\rm MNA}+V_{\rm NA}$
changes  the shape only slightly in the displayed range.
We see that,
at short distances, the FTRS result reproduces well the shape of $E_{\rm full}$ for a larger $\mu$,
while at larger distances it reproduces the shape of $E_{\rm full}$ for a smaller $\mu$.
In general, the convergence of the potentials are better at smaller $r$ both for the FTRS and
fixed-order calculations, and $r\Lambda_{\msbar}$ would be reasonably smaller than unity in
the displayed range.
Hence,
we observe a plausible feature in this comparison.

We have also examined $V_{\rm QCD}^{\rm MNA}+V_{\rm NA}$ in the DSRS method.
At $r\lesssim 0.2~{\rm GeV}^{-1}$, the DSRS result agrees well with the FTRS result, while
at larger distances it deviates from the FTRS result (and also from the fixed-order result).
In view of the analysis in Sec.~\ref{sec:2} this indicates that the convergence of the DSRS result
is quite slow compared to the FTRS result.

\section{Summary and Conclusion}
\label{sec:5}

In Sec.~\ref{sec:2} we have seen that in the LL approximation of the FTRS method $V_{\rm NA}(r)$
includes the ${\cal O}(\LQ^2/m)$
renormalon.
In Sec.~\ref{sec:3} we have seen that the fixed-order perturbative
series of the total energy of the quarkonium system improves 
stability against scale variation and convergence
as we incorporate possible cancellation of the ${\cal O}(\LQ)$ and ${\cal O}(\LQ^2/m)$
renormalons successively.
In Sec.~\ref{sec:4} we have seen that the FTRS and DSRS estimates of the ${\cal O}(\LQ^2/m)$
renormalons of 
$V_{\mathrm{NA}}$ and $2\, m_{\rm pole}^{\rm MNA}$  from the known first two terms
do not show a sign of convergence, while
there is a tendency of partial cancellation between them.
On the other hand, the FTRS estimate of  the PV part of
$V_{\rm QCD}^{\rm MNA}(r)+V_{\rm NA}(r)$ shows
a reasonable agreement with the fixed-order total energy at $r< 2.5~{\rm GeV}^{-1}$.

These analysis results are mutually consistent 
and also consistent with the hypothesis that 
both $2m_{\rm pole}$ and $V_{1/r^2}$ include the ${\cal O}(\LQ^2/m)$
renormalons and that they cancel each other.
Since there is no argument which forbids existence of the ${\cal O}(\LQ^2/m)$ renormalons,
we interpret our analysis results to be positive indications for their
existence and cancellation.

The analysis in Sec.~\ref{sec:3} indicates that already in the current status there is a
possibility to improve the accuracy of the theoretical prediction for the quarkonium
energy by incorporating cancellation of  the ${\cal O}(\LQ^2/m)$
renormalons.
Note that the current calculations of the heavy quarkonium spectrum
do not incorporate this cancellation mechanism due to the use of the $\epsilon$ expansion \cite{Hoang:1998ng}.
A proper account of this cancellation may improve accuracy of the charm and bottom quark mass
determinations from
the charmonium and bottomonium $1S$ energy levels \cite{Kiyo:2015ufa,Peset:2018ria}, for instance.

\section*{Acknowledgement}
The work of Y.S.\ was supported in part
by JSPS KAKENHI Grant Number JP23K03404.

\end{document}